# Large low field room temperature magneto-dielectric response from $(Sr_{0.5}Ba_{0.5})Nb_2O_6$ / $Co(Cr_{0.4}Fe_{1.6})O_4$ bulk 3-0 composites

**Satyapal Singh Rathore and Satish Vitta***

**Department of Metallurgical Engineering and Materials Science**

**Indian Institute of Technology Bombay**

**Mumbai 400076; India**

## Abstract

Bulk magnetodielectric composites with a 3-0 configuration comprising of ferroelectric-magnetostrictive phases have been synthesized using $(Sr_{0.5}Ba_{0.5})NbO_6$ – $Co(Cr_{0.4}Fe_{1.6})O_4$ as the two constituents respectively. The ferroelectric phase made by a dual stage sintering process has a uniform grain size of 15 μm while the magnetostrictive phase has a grain size of 2 – 3 μm. Composites synthesized by conventional solid state processing using these two constituents exhibits large magnetodielectric coupling at room temperature which increases with increasing magnetic field. The composite with 30 % magnetostrictive phase distributed uniformly in the ferroelectric phase has the most desirable microstructure and exhibits a large coupling with 3.2 % change in the dielectric constant at 1 kHz and 8 kOe magnetic field. This change in dielectric constant decreases when the fraction of magnetostrictive phase is varied, indicating that 30 % is the optimal value to realize large coupling between the two phases. This result confirms that 3D percolation causes the coupling to decrease in composites.

***For correspondence: satish.vitta@iitb.ac.in**

**Introduction:**

Materials which exhibit at least two of several types of ordering phenomena such as ferromagnetism, ferroelectricity, ferroelasticity and non-collinear magnetic spiral structures are of considerable importance as they find application in multifunctional devices.[1-3] Of these, materials that exhibit magnetic and electric order, known as magnetoelectric materials, have been investigated widely because of their use in a variety of sensor based devices.[4-8] The magnetoelectric response, although has been observed in some single phase materials, it is not of technological interest as it manifests either at very low temperatures or the coupling is very weak. This has led to the development of composite materials made of magnetostrictive and electrostrictive component materials which couple *via* strain. In these composites an external magnetic (electric) field induces strain in the ferromagnetic (ferroelectric) material due to magnetostriction (electrostriction) which in turn imposes a stress on the ferroelectric (ferromagnetic) material leading to a change in its polarization (magnetization) due to piezoelectric (magnetostrictive) behavior.[9] Neither of the materials in the composite exhibits magnetoelectric response individually and hence the collective behavior is considered a tensor product property. Since the coupling is via the strain field, the magnetoelectric response function is given as;[10]

$$\partial P / \partial H = \alpha = k_c e^m e^p \qquad (1)$$

where P is the polarization, H the external magnetic field, α the magnetoelectric coupling coefficient, $k_c$ the coupling factor, $e^m$ and $e^p$ are respectively the magnetostrictive and electrostrictive coefficients of the composite constituents. It can be seen from the above relation that the magnetoelectric coupling coefficient α is a function of both intrinsic materials parameters $e^m$ and $e^p$ and an extrinsic parameter $k_c$. Hence in composites the coupling between the two order phenomena can be maximized by;

1. Choosing constituent materials which have inherently large $e^m$ and $e^p$, and
2. Controlling the structure of the composite such that $k_c$ is large.

Various strategies wherein the effective interaction/strain transfer between the two constituents can be changed by varying the dimensionality of two constituent materials such as particulate composites, laminate composites and rod/fiber composites have been investigated to increase $k_c$ and hence $\alpha$.[11-14] A trilayer laminate composite made of Terfenol-D/PZT was found to exhibit a magneto-dielectric change of 15 % maximum at room temperature at the resonant frequency maximum of 100 kHz.[15] In the case of a nanocomposite thin film made of NZFO and PVDF, a giant magneto-dielectric change of 5.1 % maximum at room temperature for frequencies > 1 kHz has been reported.[16] In the present work the microstructure of a bulk particulate composite made of materials with large $e^m$ and $e^p$ has been optimized to obtain a large coupling at room temperature and small magnetic fields. This strategy has been chosen because it is relatively easy to synthesize particulate composites and also they are more cost effective compared to other types of composites. In the case of bulk composites, two separate configurations, 0-3 and 3-0 are possible. In the 0-3 configuration the piezoelectric phase is distributed in the matrix of piezomagnetic phase while in the 3-0 configuration an inverse distribution will be present. The magnetoelectric coupling in the two cases however is different and the 3-0 configuration has been found to exhibit a higher coupling constant.[17] Hence in the present work bulk composites with 3-0 configuration have been developed using simple synthesis techniques.

The composites comprising of Cr-substituted Co-ferrite, $Co(Cr_{0.4}Fe_{1.6})O_4$ [CCF], and Sr-Ba-Nb-oxide, $(Sr_{0.5}Ba_{0.5})NbO_6$ [SBN] with different fractions of the two constituents have been synthesized and investigated. The large magnetostrictive strain and strain derivative of CCF coupled with low electrical conductivity compared to the metallic alloy Terfenol makes it a suitable candidate. The Curie temperature for this compound has been found to be 623 K, a value much higher than room temperature and hence suitable for room temperature application. The magnetostrictive strain has been reported to be as large as 80 ppm at room temperature with the strain derivative as high as 2 nm$A^{-1}$ for ~ 20 % substitution of Fe with Cr in Co-ferrite.[18] Also, this compound is known to have a negative magnetostriction at room temperature for all external magnetic fields. The relatively low electrical conductivity minimizes dielectric loss and leads to better sensitivity. The SBN compounds exhibit relaxor like behavior with large electrostrictive coefficients. The transition temperature varies with Sr/Ba ratio in the compound between 300 K and 425 K and hence facilitates tuning it to the required temperature. The most interesting aspect of these compounds is that they exhibit a remnant polarization with large coercive fields. The electrostrictive constant in these compounds has been found to be of the order of 90 $pCN^{-1}$, a value much larger than that of conventional ferroelectric material $BaTiO_3$.[19,20] Hence in the present work SBN with a transition temperature of ~ 375 K has been chosen as the electrostrictive component in the composite. This compound has been synthesized using a modified processing technique in order to obtain a narrow, uniform grain size distribution.[21] The uniform grain size SBN is mixed with CCF in different fractions to synthesize the composites. The magnetic, dielectric and magneto-dielectric response of these composites has been investigated vis-à-vis the composite microstructure. It is found that composites with ~ 30 % magnetostrictive component exhibit the highest magneto-dielectric change at room temperature in the presence of a magnetic field.

## Experimental Methods:

The Sr-Ba-Niobate (SBN) and Cr-substituted Co-ferrite (CCF) were synthesized using conventional solid state reaction method. Stoichiometric amounts of high purity $SrCO_3$, $BaCO_3$ and $Nb_2O_5$ were used as precursors to prepare SBN. The powders were initially ground in a mortar pestle and calcined twice in air at 1300°C for 12 hrs. to ensure formation of a uniform single phase powder without any compositional variation. Post calcination, the SBN powder is mixed and ground in a centrifugal ball mill followed by green compaction. The compacted powder pellet is subjected to a two stage sintering process to obtain a single phase sintered pellet with a narrow grain size distribution. This processing results in the formation of SBN50 with a high dielectric constant compared to a single stage sintered SBN. To synthesize CCF, high purity $Cr_2O_3$, CoO and $Fe_2O_3$ were taken as starting materials. The powders were mixed and calcined at 1200°C for 12 hours to get the required phase with uniform composition. The calcined powder is then ground and green compacted before sintering at 1200°C for 3 hrs. The single phase powders, SBN and CCF, were mixed in different proportions to synthesize the composites, $(CCF)_x/(SBN)_{1-x}$ where 'x' is the mole fraction and varies from 0 to 1 (0.1, 0.3, 0.5, 0.7 and 0.9). To make the composites, the constituent powders were ball milled for 24 hrs. followed by green compaction into high density discs. The discs were then sintered in air at $1200^{\circ}C$ for 6 hrs. and these discs were used for all structural and physical properties characterization.

The structural characterization was performed by X-ray diffraction, scanning electron microscopy and energy dispersive X-ray analysis for chemical composition. The X-ray diffraction patterns obtained using Cu $K_{\alpha}$-radiation were analyzed for phase purity and crystal structure determination using FULLPROF Rietveld refinement. The magnetic properties were measured using a Vibrating Sample Magnetometer while the dielectric constant was measured using a Broadband Dielectric Spectrometer in the frequency range 1 Hz to 100 MHz. For measuring the dielectric and magneto-dielectric response the sintered high density (> 95% of theoretical density) discs were polished to generate a smooth surface and then coated with air drying Ag-paint for electrical contact. The magneto-dielectric response was measured at room temperature at 1 kHz frequency in the presence of varying external d.c. magnetic field.

**Results and Discussion**:

The results of X-ray diffraction together with that of structural refinement are shown in Figure 1 and the crystallographic parameters of the two constituents, SBN and CCF, obtained from refinement are given in Table 1(a) & (b) respectively. The SBN has a tetragonal non-centrosymmetric tungsten bronze structure belonging to the $P_4bm$ space group. The non-centrosymmetric structure with 5 formula units per unit cell can be visualized as a network of corner sharing Nb-octahedra which results in the formation of three types of interstitial sites. The smallest of the interstitial sites is totally unoccupied while the relatively larger sites are occupied by $Sr^{2+}$ and $Ba^{2+}$. Partial occupancy of the larger sites by $Ba^{2+}$ and $Sr^{2+}$ and non-occupancy of the smaller triangular interstitial sites formed by 3 corner sharing Nb-octahedra has been observed earlier[22] and they provide sites for cation hopping. The CCF on the other hand crystallizes into a cubic spinel structure belonging to $Fd\bar{3}m$ space group, Table 1(b). The non-substituted $CoFe_2O_4$ has been reported to have an inverse spinel structure with the $Fe^{3+}$ ions occupying both the tetrahedral and octahedral sites.[23] In the present case, partial substitution has been found to result in the presence of both $Cr^{2+}$ and $Cr^{3+}$ ions which occupy both tetrahedral and octahedral sites and hence result in forming a mixed spinel structure. These results clearly show that both SBN and CCF are single phase with no impurity phases and crystallize with $P4bm$ and $Fd\bar{3}m$ crystal structures respectively. The synthesis of composites using the two constituents, SBN and CCF, by solid state sintering at 1200°C can result in the formation of unwanted impurity phases. Hence, in order to determine the phase purity of the synthesized composites, the X-ray diffraction patterns of the different composites were also analyzed using two phase Rietveld refinement method. The results of structural refinement are shown in Figure 1 together with the experimental data. All the peaks in the diffractograms could be identified with Bragg peaks position corresponding to the two constituent phases in all the composites. The agreement between the experimental results and the refined structure clearly show that the two constituent phases do not react with each other during sintering. Also, the phase fractions obtained from refinement given in Table 2 are in good agreement with actual data further confirming the absence of any reaction between the two phases.

The magnetostrictive/ piezoelectric as well as the dielectric constants in composites are a strong function of the microstructure. In order to realize a large coupling constant and hence a large magneto-dielectric change it is essential that the two constituents are in effective, optimal contact with each other. The microstructure of the composites as observed in a scanning electron microscope is shown in Figure 2. The microstructure of SBN, Figure 2 (a) indicates formation of grains with a uniform size distribution and a average grain size of ~ 15 μm. The largest grains were found to be ~ 22 μm while the smallest grains are of 8 μm size. The grains have a more rounded smooth morphology not very typical of oxide materials in general. The grain size control as well as the rounded surface morphology are due to small compositional variations across the grain boundary which changes the effective liquidus temperature during sintering.[24] The morphology of CCF grains on the other hand, seen in Figure 2(f) is very different compared to the SBN grains. The grains have a highly faceted morphology with sharp features and prismatic shapes. The average grain size, in sharp contrast to the SBN grains is ~ 2 - 3 μm, an order of magnitude small. This difference in grain size between the two constituents is extremely useful as in the composite it results in efficient surface contact between the two phases, magnetostrictive and electrostrictive and also an efficient packing. The near complete absence of porosity in both the phases shows that the density is nearly equal to theoretical density. The magnetostrictive / electrostrictive coupling constant in composites is a strong function of the interphase contact and higher the contact area, higher will be the coupling efficiency. In order to delineate the grain boundaries between the two phases clearly, back scattered electron imaging was performed to study the microstructure of composites. The bright phase with large grains represents SBN and the relatively dark phase with a smaller grain size is CCF in the micrographs, Figures 2(a) to (f). The presence of just two phases clearly confirms the XRD results which show that no impurity phase or reacted phase is present in the composites. The most interesting aspect of the composite microstructures is the spatial distribution of the two phases and the resulting inter-phase coupling contact. The smaller CCF grains are randomly distributed in the matrix of large SBN grains. For concentrations ≥ 30%, the smaller CCF grains effectively cover the larger SBN grain surfaces from all sides which should lead to an highly effective coupling and hence an efficient strain transfer across the grain boundaries between the two phases. The microstructure in all the different composites is extremely dense with no porosity. This shows that the two phases are efficiently packed in spite of the large difference in their grain sizes.

The variation of magnetization M with external magnetic field H at room temperature studied both for the pure component phases as well as the composites is shown in Figure 3. The saturation magnetization $M_s$ of the ferrimagnetic component CCF is found to be 60.3 emu g$^{-1}$, in close agreement with reported bulk values.[25] The atomic magnetic moment 'm' determined from the saturation magnetization assuming 8 formula units per unit cell as per the crystal structure, is found to be ~ 2.54 $\mu_B$. This result confirms that CCF indeed has a mixed spinel structure, in agreement with the crystal structure obtained from X-ray diffraction pattern. The coercivity of CCF is found to be ~ 25 Oe, conforming to the soft magnetic nature of this ferrimagnetic phase. The saturation magnetization decreases nearly linearly with increasing SBN phase in the composites as seen in the inset of Figure 3. It is to be noted here that the magnetization saturates for fields ~ 5000 Oe indicating that magneto dielectric changes should also reach a constant value at these fields. To evaluate the contribution of CCF phase to magnetization of the composites as well as to rule out formation of any magnetic impurity phase during sintering, the isothermal magnetization was weight normalized. The results are shown in the inset of Figure 3, which shows that all the M-H curves collapse into a single loop corresponding to the pure CCF phase. These results further confirm that no additional magnetic or non-magnetic phases form during the process of making the composites.

The dielectric response of SBN, CCF and the composites with different fractions of the two phases was studied as a function of temperature from 223 K to 473 K, in the frequency range 10 Hz to 10 MHz and is shown in Figure 4(a). The phase transition in the case of SBN is diffuse, a typical characteristic of a relaxor with a paraelectric-ferroelectric transition taking place at 380 K. The dielectric constant in paraelectric state in these materials follows a modified Curie-Weiss relation given by;[26]

$$\left(\frac{1}{\varepsilon'} - \frac{1}{\varepsilon'_m}\right) = \frac{(T - T_P)^{\gamma}}{C} \qquad (2)$$

where $T_P$ is the transition temperature, C the Curie-constant and $\upsilon$ an exponent which is a measure of the nature of phase transition. In the present case the exponent $\upsilon$ is found to be ~ 1.42, a value typically observed in relaxor ferroelectrics.[27] It is to be noted here that the absolute value of the dielectric constant is > $10^3$ and remains high for all frequencies. The dielectric constant of CCF on the other hand exhibits a highly frequency dependent behavior with large values at low frequencies which decreases rapidly with increasing frequency, Figure 4(f). The dielectric constant drops from ~ $10^3$ at 10 Hz to ~ 15 at 130 kHz at 380 K, the ferroelectric transition temperature of SBN. This behavior clearly shows that contribution to the overall dielectric constant in the composite at high temperatures and high frequencies is essentially due to SBN. The dielectric relaxation is highly dispersive in nature and the peak in dielectric loss shifts to higher temperatures with increasing frequency. The magnitude of peak shift with change in frequency is quite strong indicating that the material has a superparaelectric behavior. This type of relaxation is thermally activated and has been found to be due to the presence of imperfections or hopping of cations such as $Cr^{3+}$, $Co^{2+}$ and $Fe^{3+}$ between the different sites A and B in the crystal lattice. In fact the observation of a mixed spinel crystal structure by X-ray diffraction leads us to believe that this type of behavior is due to ionic hopping between different energetically equivalent crystallographic sites. The activation energy for this ionic hopping is found to be ~ 52 meV, in agreement with earlier reported values.

The variation of dielectric constant with temperature at two different frequencies, 12 kHz and 1.3 MHz of the different composites is shown in Figures 4(b) to (e). Addition of just 10% magnetic CCF to ferroelectric SBN decreases the transition temperature to 278K from 380K. The peak becomes extremely broad and nearly vanishes for additions > 10% CCF. The dielectric constant on the other hand does not decrease significantly compared to that of pure SBN as the dielectric constant of the composites is dominated by that of SBN, which is inherently a ferroelectric material. The resistivity of CCF has been found to be about 2 orders of magnitude lower than that of SBN.[28] This results in an overall increase in conductivity of the composite due to increasing addition of CCF which results in increasing the dielectric loss significantly as seen in Figure 5. The dielectric studies of 70 % and 90 % CCF composites were also performed (not shown) and they exhibit very large dielectric loss in agreement with increase in conductivity of the composites. The dielectric relaxation in composites with widely different electrical conductivity can lead to accumulation of charge at the two phase interfaces and this leads to significant loss.

Since the ferroelectric transition temperature of SBN is ~ 380 K and the Curie temperature of CCF is 623 K, the magneto-dielectric coupling behavior of the composites has been investigated at 300 K and the results are shown in Figure 6. The magneto-dielectric response Γ defined as;[29]

$$\Gamma = \frac{\varepsilon(H) - \varepsilon(0)}{\varepsilon(0)} = (\mu_{me}/E)H \qquad (3)$$

where ε is the dielectric constant, $\mu_{me}$ the magneto-electric susceptibility, E the applied electric field and H the external magnetic field, was measured at an electric field frequency of 1 kHz as a function of varying H for different composites. The response Γ was found to be independent of the direction of magnetic field as seen in the inset of Figure 6. The dielectric constant ε decreases on application of magnetic field and the rate of change is found to be maximum at low fields, reaching a saturation with increasing H. The magneto-dielectric response was found to increase with increasing CCF content till 30 % and then decrease with increasing CCF content. The magnetodielectric response reached a maximum of ~ 3.2 % for the composite with 30 % CCF at room temperature and 1 kHz electric field frequency in the presence of 6 kOe magnetic field. Theoretically, it is predicted that the magnetoelectric coupling increases with increasing content of the magnetic phase and reaches a maximum for a volume fraction of ~ 0.5. However, the magnetic phase which generally has a higher electrical conductivity compared to the piezoelectric phase, leads to percolation conduction at these large concentrations. This leads to large dielectric losses as well as decreased magnetoelectric coupling. Hence, to circumvent this phenomenon, the concentration of the conducting magnetic phase should be lower than the percolation limit for 3D materials.[30] The results obtained in the present SBN/CCF composites are in complete agreement with these predictions. The coupling response Γ increases with increasing CCF content up to 30 % and then decreases, clearly indicating that percolation dominates at higher fractions. Also, the microstructure in the 30 % CCF composite provides optimized connectivity between SBN and CCF grains to maximize the strain transfer. This leads to large magnetodielectric response in these composites.

**Conclusions**:

A bulk magnetodielectric composite with Sr-Ba-niobate and Cr-substituted Co-ferrite as the ferroelectric and magnetostrictive phases has been synthesized using simple, conventional solid state techniques. The uniform grain size in the two phases coupled with an order of magnitude difference in their size leads to an efficient packing as well as a highly effective strain transfer contact between the two phases. This results in producing a large magnetodielectric change at room temperature in the presence of an external magnetic field. The magnitude of change in dielectric constant is found to be ~ 3.2 %, a large value for a bulk composite synthesized using standard techniques. For a 3-0 composite wherein the magnetostrictive phase is uniformly distributed around the ferroelectric phase, the volume fraction of the magnetic, conducting phase should be below the conduction percolation limit to realize maximum magnetoelectric coupling.

**Acknowledgements:** The authors acknowledge provision of access to Central Facilities of IIT Bombay.

**Table 1 (a).** The fractional position of different ions in the compound $(Sr_{0.5}Ba_{0.5})Nb_2O_6$ obtained from Rietveld refinement of the x-ray diffraction pattern at 300 K. The refined lattice parameters are; a = b = 12.445(1) Å and c = 3.9397(5) Å, α = β = γ = 90$^0$.

| Site | Wyckoff | x | y | z | Occupation |
|------|---------|---|---|---|------------|
| Nb1 | 2b | 0 | 0.5 | 0.2209(3) | 0.25 |
| Nb2 | 8d | 0.0753(2) | 0.2105(2) | 0.2140(4) | 1 |
| Sr1 | 2 a | 0 | 0 | 0.6609(6) | 0.25 |
| Sr2 | 4c | 0.1748(2) | 0.6748(2) | 0.6875(4) | 0.125 |
| Ba | 4c | 0.1748(2) | 0.6748(2) | 0.6875(4) | 0.375 |
| O1 | 8d | 0.3371(2) | 0.0137(8) | 0.2525(2) | 1 |
| O2 | 8d | 0.1495(9) | 0.0647(9) | 0.2395(7) | 1 |
| O3 | 4c | 0.2796(8) | 0.77968 | 0.2695(3) | 0.5 |
| O4 | 2b | 0.5 | 0 | 0.4788(9) | 0.25 |
| O5 | 8 d | 0.5680(8) | 0.2679(9) | 0.7001(7) | 1 |

**Table 1 (b).** The different fractional atomic positions and occupation probability at different sites in the compound $CoCr_{0.4}Fe_{1.6}O_4$ used for Rietveld refinement of the room temperature XRD. The lattice parameters obtained from refinement are; a = b = c = 8.3831(7) Å, $\alpha = \beta = \gamma = 90^0$.

| Atom | Wyckoff | X | Y | Z | Occupation |
|------|---------|-------|-------|-------|------------|
| Co1 | 8a | 0.12500 | 0.12500 | 0.12500 | 0.03952 |
| Fe1 | 8a | 0.12500 | 0.12500 | 0.12500 | 0.00208 |
| Fe2 | 16d | 0.500 | 0.500 | 0.500 | 0.06254 |
| Co2 | 16d | 0.500 | 0.500 | 0.500 | 0.00416 |
| Cr1 | 16d | 0.500 | 0.500 | 0.500 | 0.01660 |
| O1 | 32e | 0.25298 | 0.25298 | 0.25298 | 0.1666 |

**Table 2.** The XRD spectra of composites was analyzed using Rietveld refinement to determine the individual phase fractions. The results are given below where $X^2$ represents the fitting quality parameter.

| S.No. | Composition | $X^2$ | Phase Fraction, SBN50/CCF |
|-------|-------------|-------|---------------------------|
| 1. | SBN50 | 2.34 | 100 |
| 2. | Comp 10 % | 2.61 | 89.28(0.37)/10.72(0.04) |
| 3. | Comp 30 % | 2.46 | 71.01(0.55)/28.99(0.06) |
| 4. | Comp 50 % | 2.37 | 51.75(0.5)/48.25(0.1) |
| 5. | Comp 70 % | 1.62 | 31.35(0.62)/68.65(0.2) |
| 6. | Comp 90 % | 1.6 | 8.91(0.74)/91.09(0.47) |
| 7. | CCF | 1.33 | 100 |

**Figure Captions:**

**Figure 1.** The Rietveld refined curve together with experimental data of XRD pattern for $(Sr_{0.5}Ba_{0.5}Nb_2O_6)_{1-x}/(CoCr_{0.4}Fe_{1.6}O_4)_x$ composites with x=0 (**a**), x = 0.3(**b**), x= 0.5(**c**), x=0.7(**d**), x=0.9(**e**), and x=1(**f**). All peaks are identified to constituent phases of the composites. The upper and lower vertical bars indicate Bragg reflections corresponding to SBN50 and CCF phase respectively.

**Figure 2.** The backscattered electron images (a to f) of $SBN_{1-x}/CCF_x$ composites for x=0 (**a**), x=0.3(**b**), x=0.5(**c**), x=0.7(**d**), x=0.9(**e**) x= 1 (**f**). The CCF phase can be seen as smaller dark grains while the SBN50 phase as larger bright grains.

**Figure 3.** The variation of magnetization M with applied magnetic field H at room temperature for $(Sr_{0.5}Ba_{0.5}Nb_2O_6)_{1-x}/(CoCr_{0.4}Fe_{1.6}O_4)_x$ composites. The number after Comp indicates mol. % of CCF phase in the composites. The upper inset shows weight normalized magnetization of the composites and the lower inset shows the variation of saturation magnetization with CCF content.

**Figure 4.** Thermal variation of dielectric constant, $\epsilon'$ for polycrystalline $(Sr_{0.5}Ba_{0.5}Nb_2O_6)_{1-x}/(CoCr_{0.4}Fe_{1.6}O_4)_x$ composites at 1.17 kHz, 12 kHz and 1.3 MHz frequencies with (**a**) x=0, (**b**) x=0.1, (**c**) x=0.3, (**d**) x=0.5 and (**e**) x=1. The transition temperature decreases with increasing CCF content in the composite.

**Figure 5.** Temperature dependence of dielectric loss tanδ for polycrystalline $(Sr_{0.5}Ba_{0.5}Nb_2O_6)_{1-x}/(CoCr_{0.4}Fe_{1.6}O_4)_x$ composites at 1.17 kHz, 12 kHz and 1.3 MHz with (**a**) x=0, (**b**) x=0.1, (**c**) x=0.3, (**d**) x=0.5 and (**e**) x=1. The dielectric loss increases with increasing CCF content.

**Figure 6.** The variation of magneto-dielectric response Γ with applied dc magnetic field at 1 kHz (electric field frequency) for Comp 10 %, Comp 30 % and Comp 50 % composites at room temperature. The dielectric response increases with increasing CCF content up to 30 % beyond which it decreases. The inset shows that the dielectric response is independent of magnetic field direction.

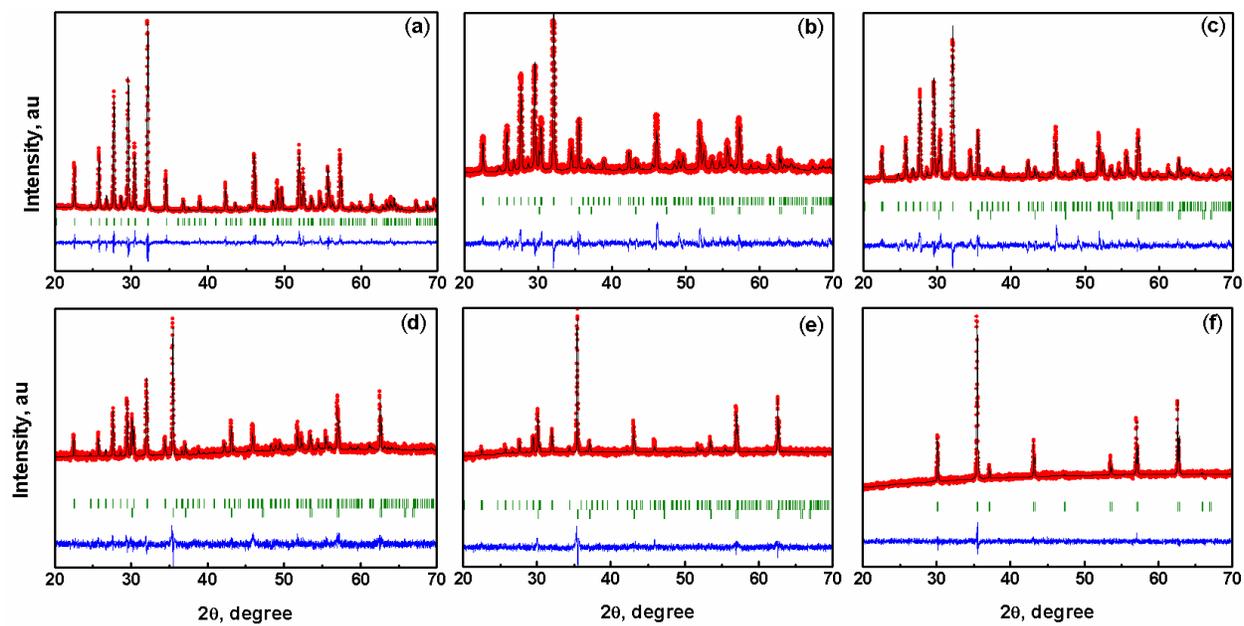

Figure 1

Figure 2

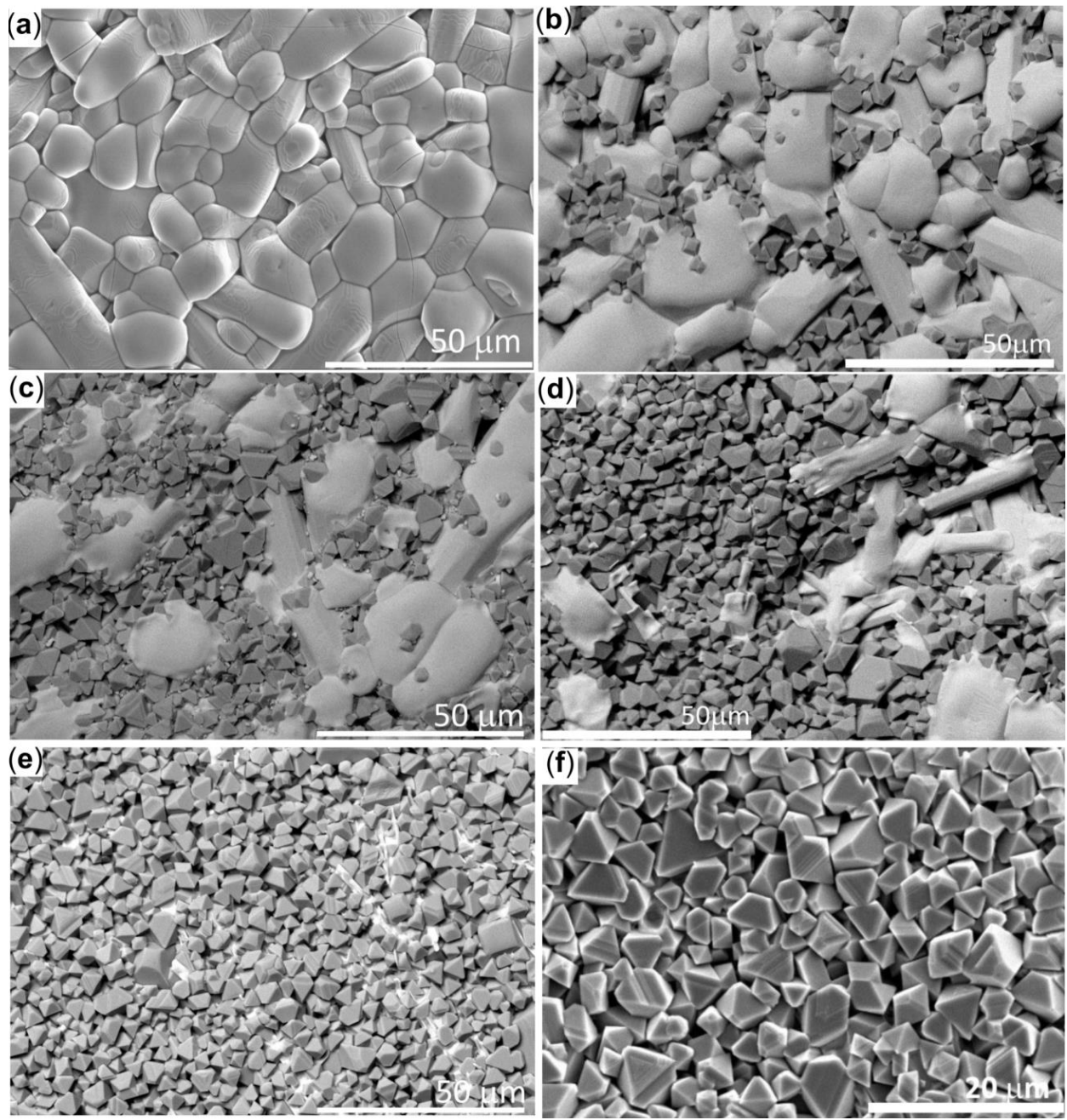

Figure 3

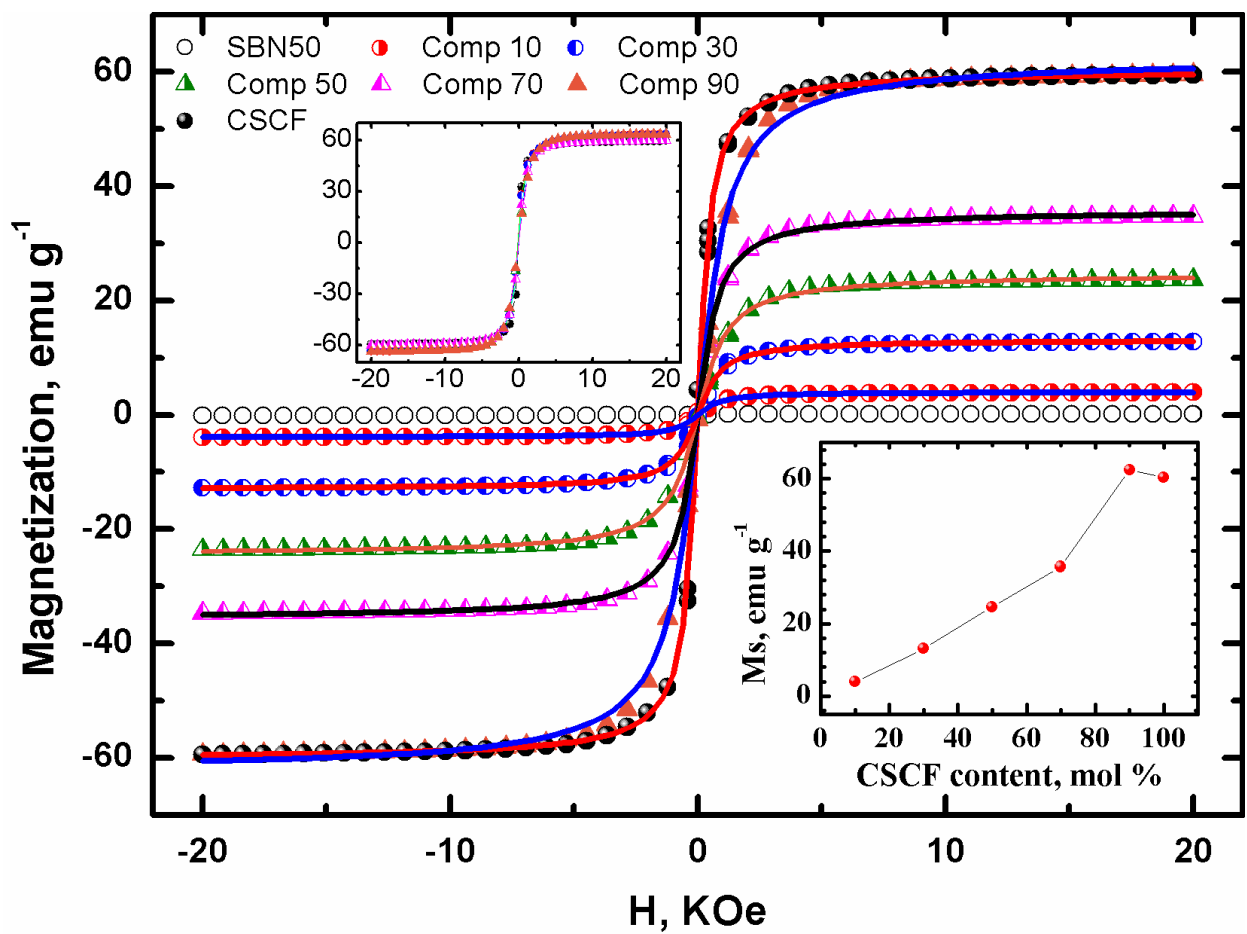

Figure 4

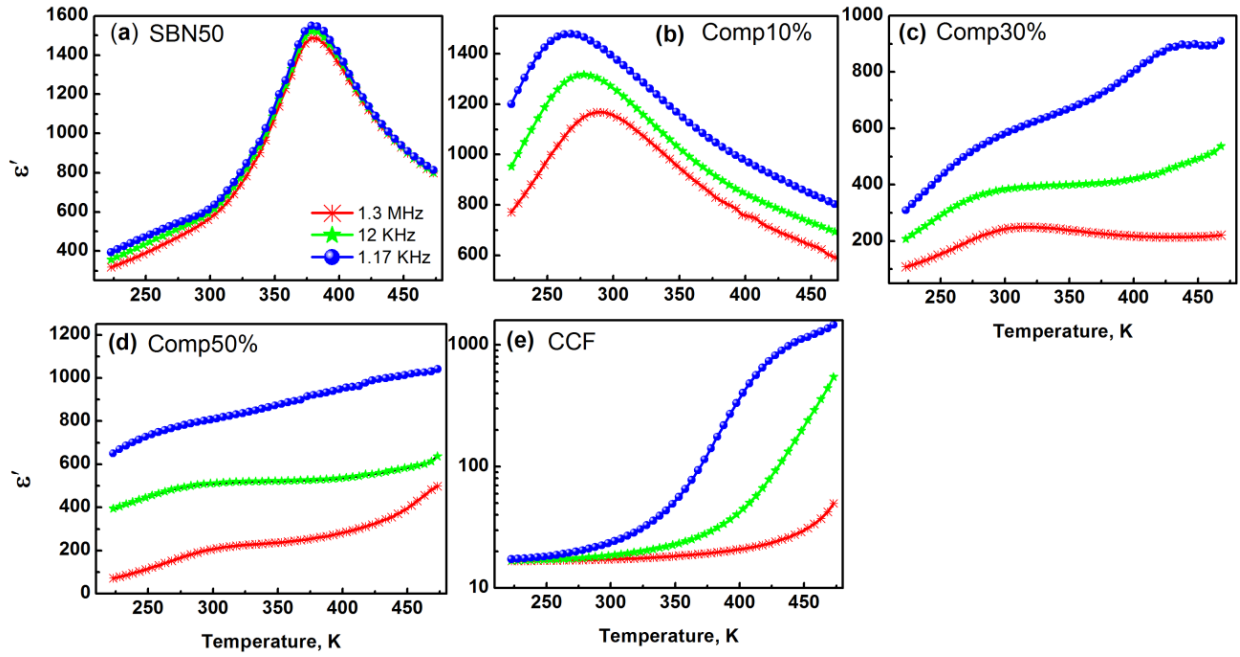

Figure 5

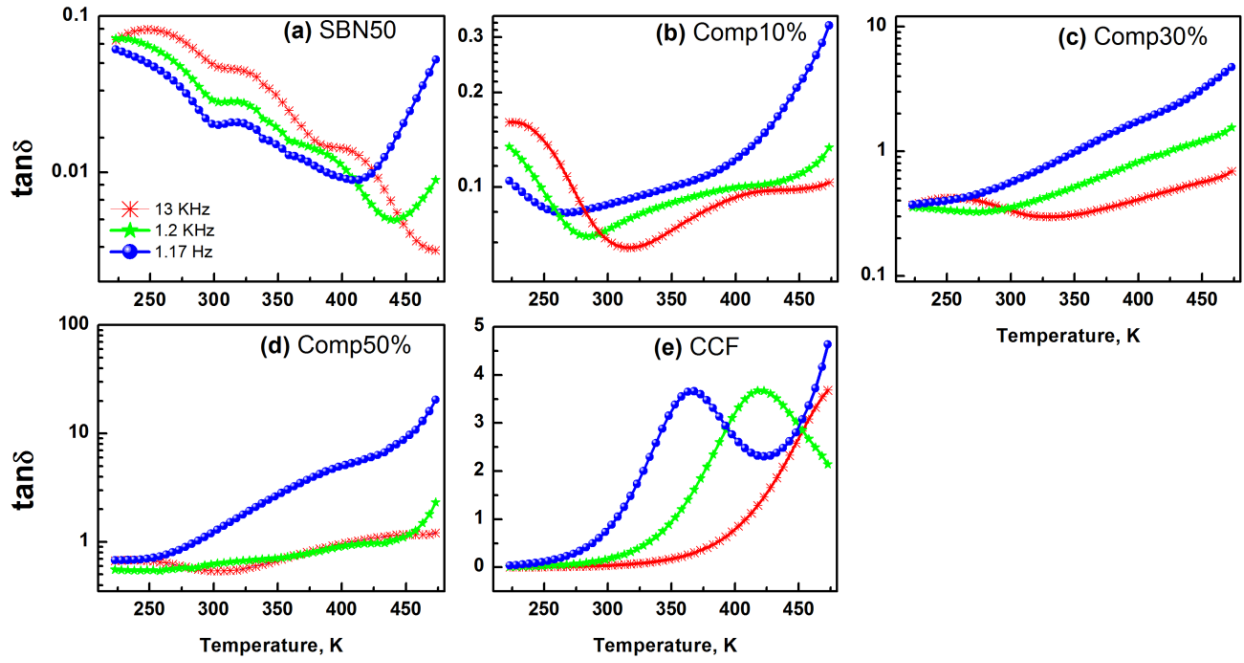

Figure 6

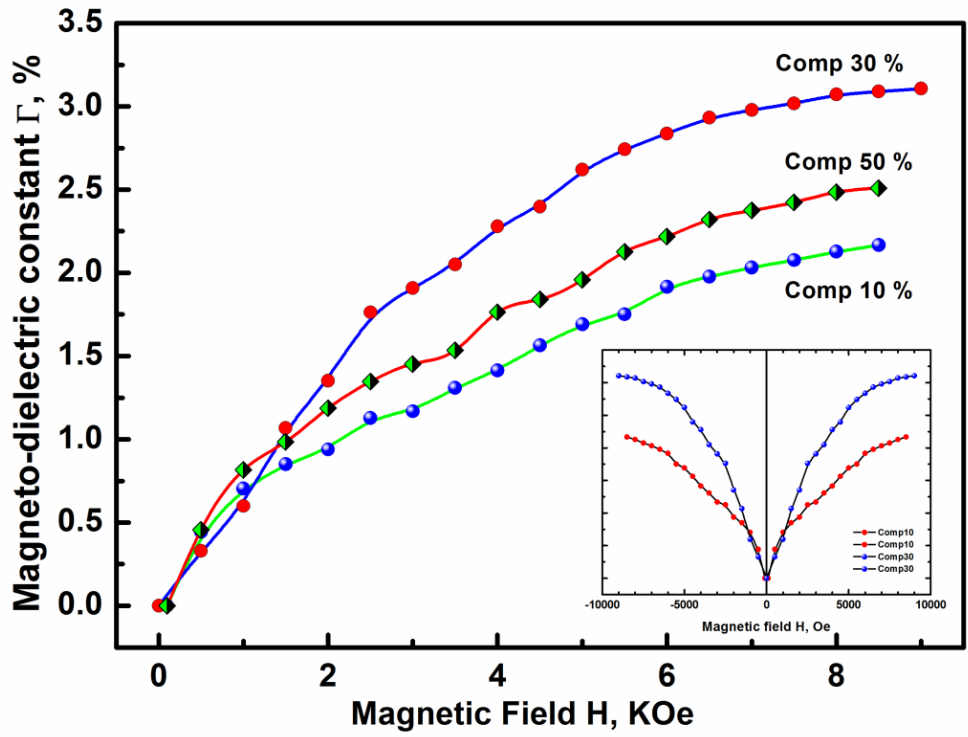